\documentclass{emulateapj}
\usepackage{graphicx,amsmath,natbib,bm}

\usepackage{epsf}
\usepackage{epstopdf}

\input{epsf}
\usepackage{epsfig}

\usepackage{amsfonts}
\DeclareGraphicsExtensions{.jpg,.pdf,.png,.eps,.ps}

\slugcomment{}

\shorttitle{Lensing Noise in SZ Surveys}
\shortauthors{Hezaveh et al.}

\begin{document}

\title{Lensing Noise in mm-wave Galaxy Cluster Surveys}
 			
\author{Yashar Hezaveh, Keith Vanderlinde, Gilbert Holder, Tijmen de Haan}
\affil{Department of Physics, McGill University,
    Montreal, QC}

\begin{abstract}
We study the effects of gravitational lensing by galaxy clusters of the background of
dusty star-forming galaxies (DSFGs) and the Cosmic Microwave Background (CMB),
and examine the implications for Sunyaev-Zel'dovich-based (SZ) 
galaxy cluster surveys.
At the locations of galaxy clusters, 
gravitational lensing modifies the probability distribution of the background flux of the  DSFGs as well as the CMB. We find that, in the case of a
single-frequency 150 GHz survey, lensing of DSFGs
leads to both a slight {\em increase} ($\sim10\%$) in detected
cluster number counts (due to a $\sim 50$\% increase in the variance of the DSFG background,
and hence an increased Eddington bias), as well as to a rare 
(occurring in $\sim2\%$ of clusters) ``filling-in'' of SZ cluster signals
by bright strongly lensed background sources. Lensing of the CMB
leads to a $\sim55\%$ reduction in CMB power at the location of massive 
galaxy clusters in a spatially-matched single-frequency filter, leading to
a net {\em decrease} in detected cluster number counts. 
We find that the increase in DSFG power and decrease in CMB power due
to lensing at cluster locations largely cancel, such that 
the net effect on cluster number counts for current SZ surveys is
sub-dominant to Poisson errors.

\end{abstract}


\section{Introduction}

Galaxy clusters are the largest collapsed objects in the universe.
In hierarchical scenarios of structure formation their abundance
strongly depends on the growth rate of structure over cosmic time,
and  measurements of their abundance as a function of mass and
redshift can impose stringent constraints on cosmological parameters \citep{Wang:98, Holder:01}.   

The majority of baryonic mass in clusters is in the hot intra-cluster medium.
In addition to emitting X-rays, this plasma interacts with Cosmic Microwave Background (CMB) photons
via inverse Compton scattering, causing a local spectral distortion in the CMB \citep{Sunyaev:72},
known as the Sunyaev-Zeldovich (SZ) effect.
This effect can be used to detect clusters in large area CMB surveys.
The selection of clusters based on this method is particularly valuable
since the surface brightness of the SZ effect is redshift-independent,
and can therefore be used to explore a large range of redshifts.
In addition the SZ signature depends strongly on the total mass of the 
clusters with small scatter \citep{Barbosa:96, Shaw:08},
allowing surveys to construct nearly mass-limited catalogs.

Using data from the South Pole Telescope (SPT) \citep{Carlstrom:11},
the first discovery of galaxy clusters via the SZ effect was reported in \citet{Staniszewski:09},
followed by the first cosmological constraints from a catalog of SZ-detected clusters in \citet{Vanderlinde:10}.
More recently \citet{Marriage:11} reported clusters discovered by the
Atacama Cosmology Telescope (ACT), with cosmological results based on that
catalog in \citet{Sehgal:11}, 
the Planck collaboration released a sample of SZ-selected
galaxy clusters \citep{Planck} and \citet{Reichardt:12} published a larger catalog from multi-frequency SPT data.
  
In SZ surveys, dusty star-forming galaxies (DSFGs),
seen as an unresolved background of point sources,
are a source of astronomical confusion which acts as a noise term in the SZ-detection of galaxy clusters;
the number counts \citep{Vieira:11} and power spectrum \citep{Reichardt:11} of these sources are well-studied.
The other main source of confusion noise is the CMB, a Gaussian field with an extremely well-characterized spatial power
spectrum \citep{Keisler:11, Das:11} on the scales of interest.

At the location of galaxy clusters, due to the efficiency of clusters as  gravitational lenses,
these backgrounds can be strongly distorted, as emphasized by \citet{Blain:98}.
These distortions can significantly affect the probability 
distributions of these sky noise backgrounds,
which can in turn affect the SZ detection statistics of such surveys
and the resulting cosmological parameter constraints.
For example, \citet{Blain:98} found that the confusion noise from
point sources can be amplified by a factor of three within an aperture
comparable to the inner region of the cluster
where strong lensing (multiple images) can occur.
In this paper we simulate gravitational lensing of background DSFGs and the CMB by galaxy clusters,
measure the probability distributions of the lensed fields,
and estimate the effects of this lensing on SZ cluster surveys. 

In section \S\ref{sec:analytical} we develop an analytical analysis of the
non-Gaussianity of the background confused galaxies and the CMB due to cluster lensing,
focusing on understanding the underlying processes that lead to this effect.
In section \S\ref{sec:simulations} we describe the ray-tracing simulations that
are performed to measure the shape of the background probability distributions
for realistic cluster mass profiles and backgrounds.
In section \S\ref{sec:SZeffect} we study the impact of this background lensing
on SZ cluster surveys, specifically a single-band SPT-like survey.
Discussion and conclusion are presented in section \S\ref{sec:discussion}.
In what follows we assume a spatially flat universe described
by the best-fit WMAP7 \citep{Komatsu:11} 
cosmological parameters ($h=0.71$, $\sigma_8=0.801, \Omega_M =  0.2669$).

\newpage

\section{Analytical analysis of background fluctuations}
\label{sec:analytical}

\subsection{Uniform Magnification}
\label{sec:heuristic}
The simplest example of background lensing
is the case of uniform isotropic magnification of the CMB by a 
small factor
of $\mu_\circ$ (close to 1) in each direction of a region of
sky. As shown in \citet{Bucher:12}, such a uniform magnification leads to
a shift in the power spectrum 
$\delta c_{\ell}/c_{\ell} = \mu_\circ(d ln c_{\ell}/d ln \ell +2)$ for small
$\mu_\circ$; for larger magnifications and a power-law power spectrum
$c_{\ell} \propto \ell^\alpha$ the 
variance will be scaled by $\mu_\circ^{2+\alpha}$. For
constant power per $ln$ interval ($c_{\ell} \propto l^{-2}$), 
there is therefore no shift, but 
one could expect a substantial difference for either the damping tail of the 
CMB, where the spectrum is substantially steeper than $\ell^{-2}$, or for a 
Poisson distribution of sources where $c_{\ell}$ is constant.

For a Poisson distribution of sources, the 
power would scale as 
$\sim \mu_\circ^2$, while in the CMB damping tail, where the 
power-law can be steeper than $l^{-4}$, the power could
scale more steeply than $\sim \mu_\circ^{-2}$. These effects therefore go in opposite
directions and are likely to be comparable in amplitude. 

\subsection{Discrete sources}

An analytical model for the flux distribution inside an aperture due to point sources can be developed
for simple cases of lensing, as well as in the absence of lensing.
This model is helpful in that it provides a test of the numerical 
implementation as well as providing some intuition for the processes that
influence the shape of the final distributions.

In this section we assume the simplest conditions: a top-hat filter, and only consider 
sources at a single redshift of $z_s=2.0$.
The differential number counts $dN/dS$ are assumed to follow the distribution
predicted by the Durham semi-analytic model \citep{Baugh:06},
but with all sources placed at a single $z$.

\subsubsection{Unlensed DSFG Background}
In the absence of lensing the number of point sources with flux between
$S_i$ and $S_i + \Delta S$ falling inside an aperture of angular radius
$\theta_{ap}$ is simply proportional to the number counts $dN/dS(S_i)$ 
and the aperture area $\pi \theta_{ap}^2$: 
\begin{equation}
<N_i> = \left[\frac{dN}{dS}(S_i)
 \Delta S \right] \; (\pi \theta_{ap}^2)
\end{equation}

To get the flux distribution we first 
consider only sources with flux $S_i$.
The distribution of the number of these sources is given by a Poisson
distribution with a mean of $<N_i>$. The probability of observing $k$ sources
each of flux $S_i$ in a given area of sky is then 
\begin{equation}
P_i(k) = \frac{<N_i>^k}{k!} \; e^{-<N_i>}  \ .
\end{equation}
Since each source has a flux of $S_i$, in terms of the total aperture flux 
$S=k S_i$ due to sources with flux $S_i$, we can write  the probability
distribution (for discrete values of $S/S_i$):
\begin{equation}
P_i(S) = \frac{<N_i>^{(S/S_i)}}{(S/S_i)!} \; e^{-<N_i>} \ .
\end{equation}

The probability distribution of the addition of two such source
populations (with different fluxes) is given by the convolution of the probabilities
$P_{i,j}(S) = (P_i * P_j) (S)$ and if all $i=0,...N$ flux bins are included the total probability is
\begin{equation}
P_{Total}(S) = (P_0 * P_1 * ... * P_N) (S)
\end{equation}

\subsubsection{Lensed DSFG Background}

In the presence of lensing, different regions of sky undergo different 
magnification factors, slightly complicating the analysis. 
For simplicity, here we consider an SIS halo \citep{Kormann:94} lensing unresolved 
point sources before moving onto detailed numerical simulations using a 
more complex lens and source model. 

If the aperture is centered on the halo, we can break the aperture area into M circular rings with equal widths $\Delta \theta$ such that
\begin{equation}
\hat{\Omega}_{ap} = \sum_{j=1}^{M} \; \hat{\Omega}_j \; = \sum_{j=1}^{M} \; 2\pi \theta_j \; \Delta \theta
\end{equation}
where $\hat{\Omega}_{ap}$ is the total area of the top-hat aperture and 
$\theta_j$ is the angular radius of each bin. 

For SIS halos with Einstein radius $\theta_E$ the magnification at position $\theta_j$ is given analytically by
\begin{equation}
\mu(\theta_j)=\frac{|\theta_j|}{|\theta_j|-\theta_{E}}
\end{equation}

We can calculate the flux distribution inside such a circular ring $j$. 
The lensed flux of each population $i$ is now $\hat{S}_{i,j}=\mu_j S_i$ 
(where the hat refers to lensed quantities). 
Lensing also changes the observed solid angles, diluting the source number counts; 
objects inside this ring truly reside in a region of the sky with 
area $\Omega_j=\hat{\Omega}_j/\mu_j$ in the absence of lensing. Therefore the number of sources in the $j$'th ring is

\begin{equation}
<\hat{N}_{i,j}> = \left[\frac{dN}{dS}(S_i)
\Delta S_i\right] \; \frac{2\pi \theta_j \; \Delta \theta_j}{\mu}
\end{equation}

The distribution of the total lensed flux ($\hat{S}=k \mu_j S_i$) coming from each source population is again 
controlled by the Poisson distribution with a mean of $<\hat{N}_{i,j}>$ 
(for discrete values of $\hat{S}/\mu_j S_i$)
\begin{equation}
P_{i,j}(\hat{S}) = \frac{<\hat{N}_{i,j}>^{(\hat{S}/(\mu_j S_i))}}{(\hat{S}/(\mu_j S_i))!} \; e^{-<\hat{N}{i,j}>}
\end{equation}

The distribution of total flux inside the $j$'th ring is 
\begin{equation}
P_{j,Tot}(\hat{S}) = (P_{i=1,j} * P_{i=2,j} * ... * P_{i=N,j}) (\hat{S})
\end{equation}
However since all the light inside the aperture is combined, 
the distribution of the total flux inside the aperture is again a 
convolution of the probabilities inside each ring
\begin{eqnarray}
P_{Total}(\hat{S}) = \left[P_{i=1, j=1} * ...* P_{i=N, j=1}\right]*  \nonumber \\
 ...*\left[P_{i=1, j=M} * ... * P_{i=N, j=M}\right] (\hat{S})
\end{eqnarray}

\begin{figure}[h]
\centering
\epsscale{1.0}
\plotone{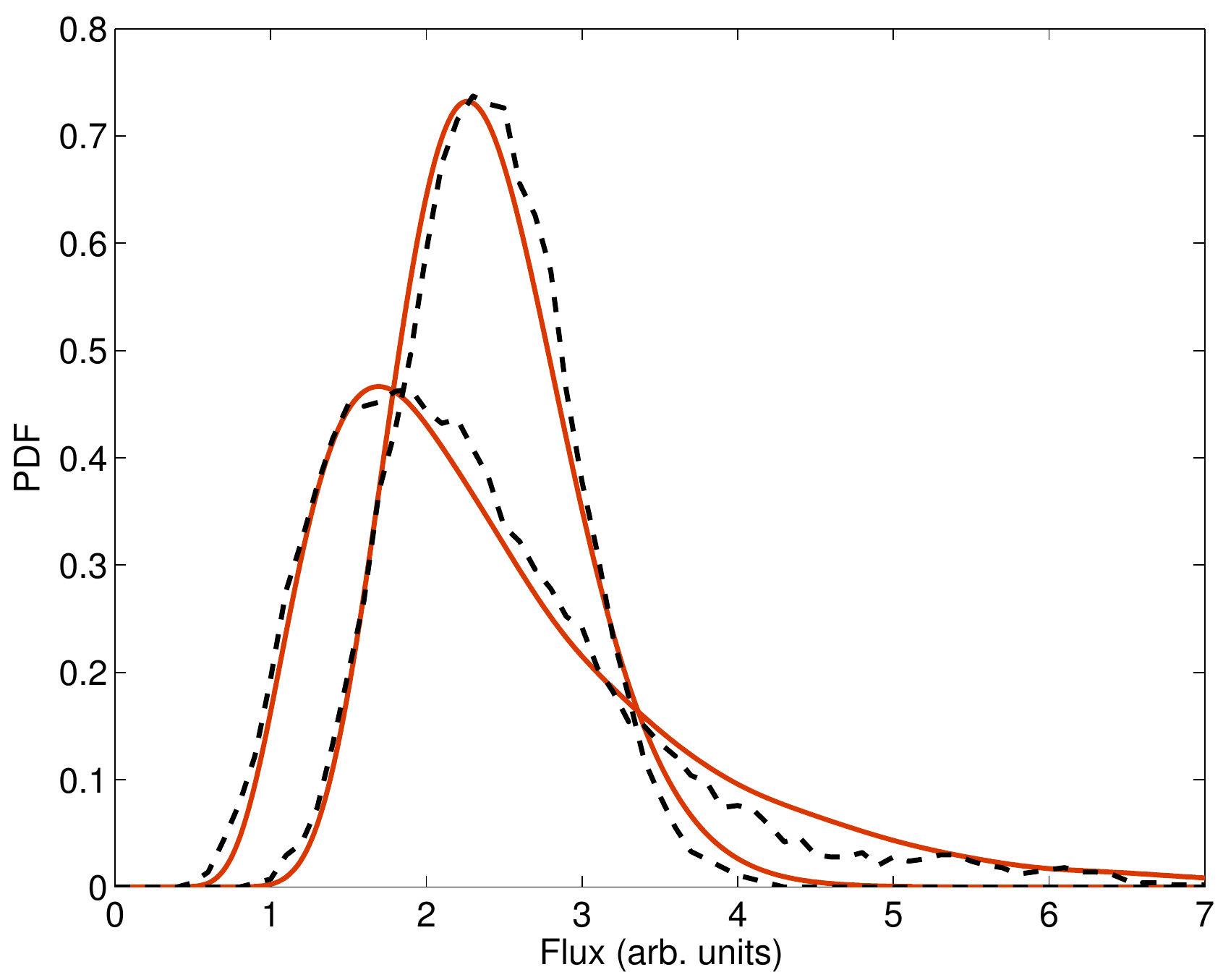}
\caption{Flux distribution inside a top-hat filter with radius $R_{filter}=0.5'$ 
centered on a circular SIS lens halo with M=$10^{15}\, M_{\odot}$ from
point sources at redshift $z_s\sim 2.0$.
The Gaussian and the skewed curves show the unlensed and the lensed flux distributions respectively.
The solid (red) curves show the analytical predictions and the
dashed (black) curves show the results of simulations.}
\label{analytical}
\end{figure}

The combined flux is shown in Figure \ref{analytical} compared with a
numerical simulation of this same configuration, showing excellent agreement.
The main effects of lensing are to shift the peak of the PDF to lower
fluxes, to broaden the distribution and to populate a high flux tail.
The peak of the distribution has moved to 
lower fluxes as there is a high probability of magnifying a spot on the sky 
that does not host a source; the distribution has broadened because there
is now more shot noise because the effective area being sampled is smaller with lower source number density;
and the high flux tail has been populated by strong lensing events where
a source happens to fall in a region of high magnification. Lensing leaves the mean of the distribution unaltered.

This calculation can be generalized to include other filters by weighting 
each region's distribution appropriately. 
However since $\mu(\theta)$ is generally not analytic for more 
complicated lens profiles, a numerical ray-tracing approach is preferred. 
In addition the above treatment assumes point sources, while in reality the 
non-zero size of sources has a substantial effect on lensing statistics
\citep{Hezaveh:12}.  For these reasons, we now turn to numerical simulations
based on ray-tracing.

\section{Simulations}
\label{sec:simulations}
We use the ray-tracing code of \citet{Hezaveh:11} to simulate realistic maps of the lensed background sources. 
As a lens model, we assume that the halo density profile in the outer regions
follows a spherical NFW profile \citep{Navarro:97} and include an SIS profile at the
halo center to mimic the enhanced baryon density at the centers of galaxy
clusters. We assume a mass-concentration relation from \citet{Mandelbaum:08} and
a central velocity dispersion of $250 \, km/s$ for the SIS component.

We generate lensing deflection maps for multiple source redshifts
(ranging from 0.7 to 4.5 for DSFGs, and one at $z=1100$ for the CMB).
For DSFGs we populate the source plane with circular galaxies of radius 1 kpc,
with number counts following the Durham model \citep{Baugh:06}. 
For each source flux bin we include $N$ sources of the given flux in the source plane map,
where $N$ is picked randomly from a Poisson distribution with a mean of $(dN/dS/dA) \,\delta S \, A_{map}$
where $\delta S$ is the width of the flux bins and $A_{map}$ is the total area of the source plane grid.
For lensing of the CMB we make realizations of random
CMB fields with power spectra calculated using CAMB \citep{Lewis:99}.

\begin{figure}[h]
\centering
\plotone{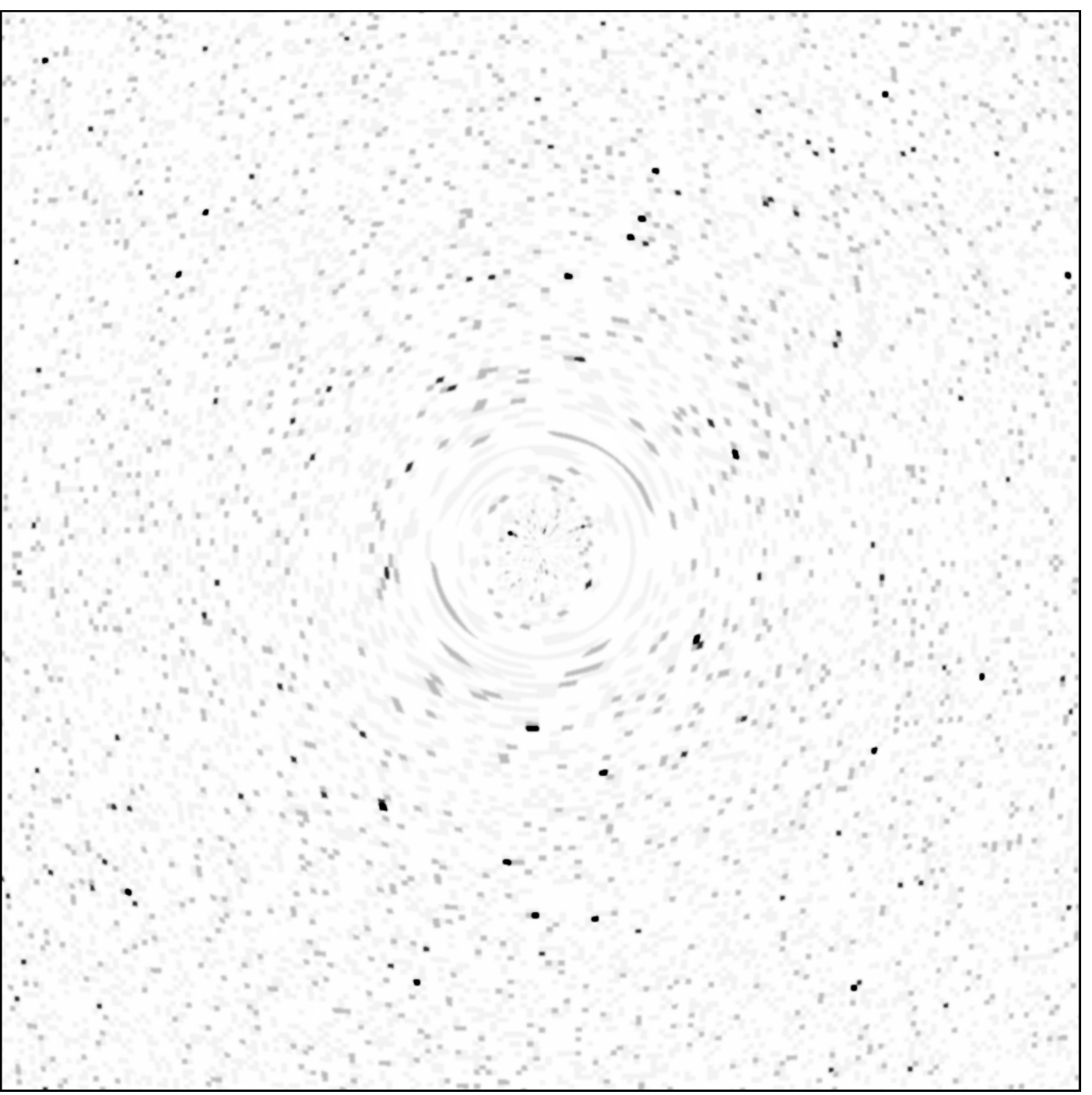}
\caption{The inner $5\times5$ arcmin of a simulated field of DSFGs for a lens of $1\times 10^{15} M_{\odot}$ at $z_d=0.5$. As it can be seen from the figure lensing of DSFGs can both magnify the sources and the empty regions of the sky.}
\label{f3}
\end{figure}

We thus generate and lens separate maps of the background DSFGs and the CMB.
An example of a lensed DSFG map is shown in Figure \ref{f3}.
It is clear from this figure that lensing can amplify the flux of a few highly magnified DSFGs while diluting their source density.

\section{Effects on SZ Cluster Surveys}
\label{sec:SZeffect}
The impact of lensing on SZ cluster surveys will depend on the details of
the experiment. Multifrequency surveys can isolate the SZ signal spectrally,
while an instrument with a large beam will be mainly sensitive to regions
of the cluster with relatively low magnification.
As a concrete example, we will consider a single-band 150 GHz
cluster survey modelled after that of \citet{Vanderlinde:10} to
investigate the impact of lensing.

\subsection{Matched-filter}
\label{sec:matchedfilter}
SZ surveys have generally employed a matched-filter approach to identifying
galaxy clusters.
The raw microwave sky maps are filtered to optimize detection
of objects with morphologies  similar to the SZ signatures expected
from galaxy clusters, through the application of spatial matched-filters 
\citep{haehnelt:96,herranz:02a,herranz:02b,melin:06}.
For the single-frequency case, in the spatial Fourier domain the map is multiplied by 
\begin{equation*}
\psi(k_x,k_y) = \frac{B(k_x,k_y) S(|\vec{k}|)}{B(k_x,k_y)^2 N_{astro}(|\vec{k}|) + N_{noise}(k_x,k_y)}
\end{equation*}
where $\psi$ is the matched-filter, $B$ is the response of the
instrument after timestream processing to signals on the sky,
$S$ is the assumed source template, and the noise power has been broken into
astrophysical ($N_{astro}$) and noise ($N_{noise}$) 
components.
For the source template, we assume
$ \Delta T \propto (1+\theta^2/\theta_c^2)^{-1}, $
where the core radius $\theta_c$ is assumed here to be 0.5',
consistent with the results found in \citet{Reichardt:12}.

This filter application is a linear operation, so the effect on the
DSFG background and the CMB 
can be considered independently.
We simulate many realizations of the DSFG field and the CMB 
(pairs of lensed and unlensed maps), apply a matched-filter similar to that used in \cite{Vanderlinde:10} to them, and record the values of the filtered maps at the center of the cluster.

\begin{figure}[h]
\centering
\epsscale{1.1}
\plotone{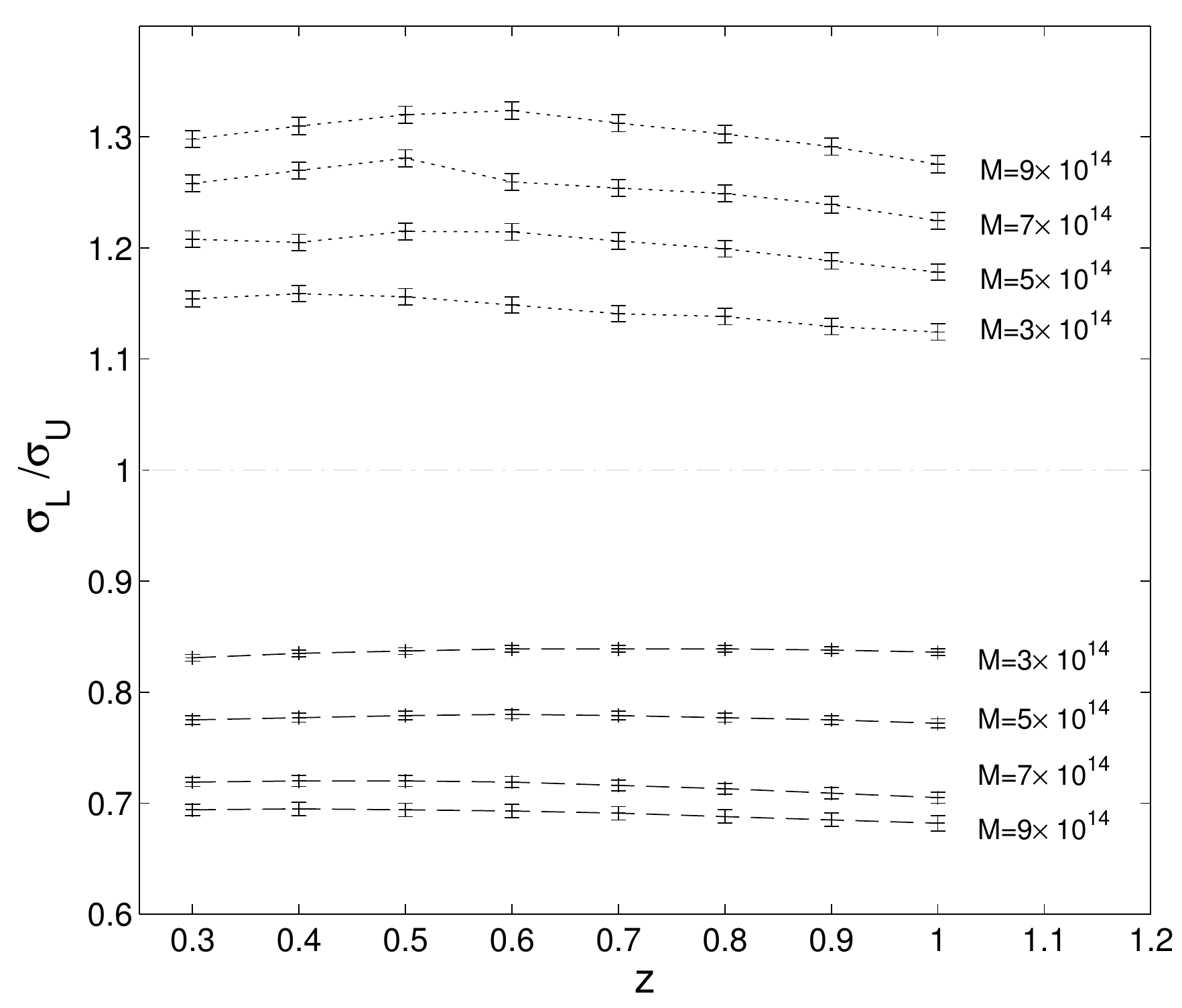}
\caption{The ratios of the standard deviations of the lensed and unlensed fields at the center of the cluster after convolution with the matched-filter. For each mass and redshift combination many different background realizations are simulated. The points connected with doted lines correspond to the DSFG background while the points connected with dashed lines correspond to CMB maps. The masses ($M_{200}$) are reported in $M_{\odot}$. } 
\label{sigma_ratio}
\end{figure}

We then fit Gaussian functions to the resulting probability distributions of the 
lensed and unlensed backgrounds. 
Figure \ref{sigma_ratio} shows the ratios of the standard deviations of the best Gaussian fits to lensed and unlensed distributions ($\sigma_L/\sigma_U$). 
This process is repeated for a variety of lens properties, stepping through mass and redshift. 

\subsection{Lensed flux distributions}

\begin{figure}[h]
\centering
\epsscale{1.2}
\plotone{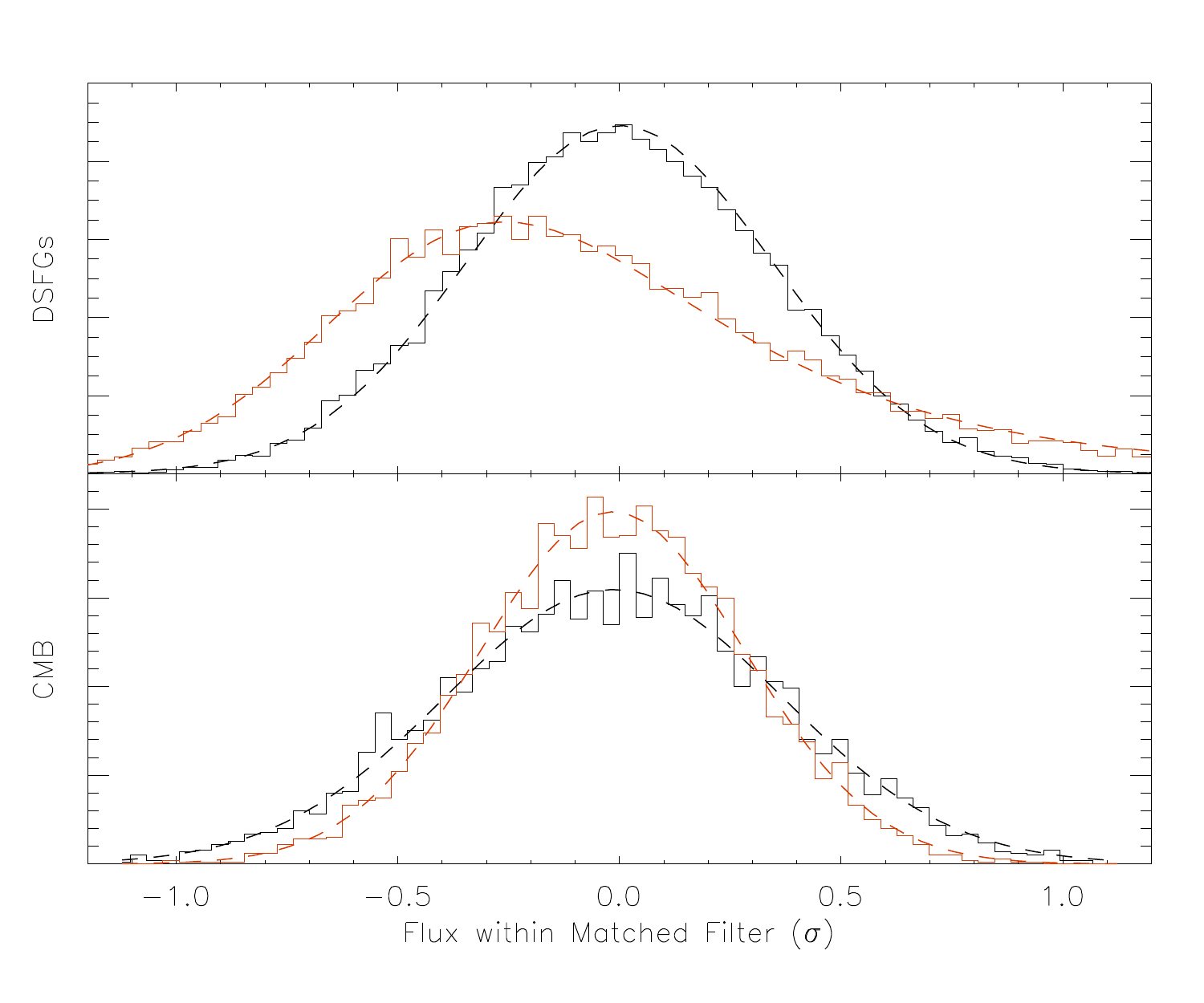}
\caption{Flux distribution of the filtered maps at the location of the cluster, in units of matched-filtered map noise.
The top panel shows the flux distribution of DSFGs while the bottom panel shows the distribution of CMB flux in the maps ($M_{200}=9\times 10^{14} M_{\odot}, \,z=0.4$). In both panels black curves correspond to unlensed and red curves to lensed distributions. The dashed curves show the Gaussian (bottom) or log-logistic (top lensed) fits to the distributions.}
\label{pdfs}
\end{figure}

Examples of the matched-filtered DSFG and the CMB flux distributions are shown in Figure \ref{pdfs}.
The top (bottom) panel shows the broadening (narrowing) of the DSFG (CMB) flux distribution by lensing, 
as expected from the simple arguments presented earlier (\S\ref{sec:heuristic}).
The high-flux tail of the distribution is caused by occasional highly magnified
lensing of DSFGs which can substantially ``fill in'' the 150 GHz SZ decrement.
In all cases, lensing conserves the mean of the distributions, keeping the average contribution from the DSFG background and the CMB unchanged.
The lensed DSFG distributions are well-fit by the log-logistic function,
\begin{equation}
N \propto (\beta/\alpha)(F/\alpha)^{\beta-1}[1 + (F/\alpha)^\beta]^{-2}
\label{eq:loglogistic}
\end{equation}
where $N$ is the Poisson expectation value at flux level $F$,
$\alpha$ is a scale parameter, $\beta$ is a shape parameter.
These parameters are found to vary as a function of the lens properties (cluster mass and redshift).

\subsection{Cluster Number Density}
By comparing the lensed and unlensed distributions we can
estimate the effect of lensing on a cluster survey.
We compute the number of detected galaxy clusters expected in a single-band SPT-like survey,
both with and without the effects of lensing on the DSFG and CMB backgrounds.

We begin by generating a mass function, a grid of expected number densities
of galaxy clusters versus mass and redshift, following the prescription of \cite{Tinker:12}.
Using the significance-mass scaling relation of \cite{Vanderlinde:10}, we 
convert this to a grid in $\zeta-$z, where $\zeta$ is the ``unbiased significance'',
defined in that work.

We convert from this grid to an observable space ($\xi-$z,
where $\xi$ is maximum S/N of a detection across filter profiles) by
convolving in the measurement uncertainty.
Without lensing of the DSFG background and the CMB, this measurement
uncertainty has unit width (by construction, the uncertainty in this space is $1\sigma$).
Lensing has the effect of slightly increasing the size of this convolution
kernel in the case of DSFGs, and decreasing the width
in the case of CMB lensing. It is important to point out that lensing simply changes the variance and the shapes of the  probability distribution of the background emissions while keeping their mean unaltered.

\begin{figure}[h]
\centering
\epsscale{1.2}
\plotone{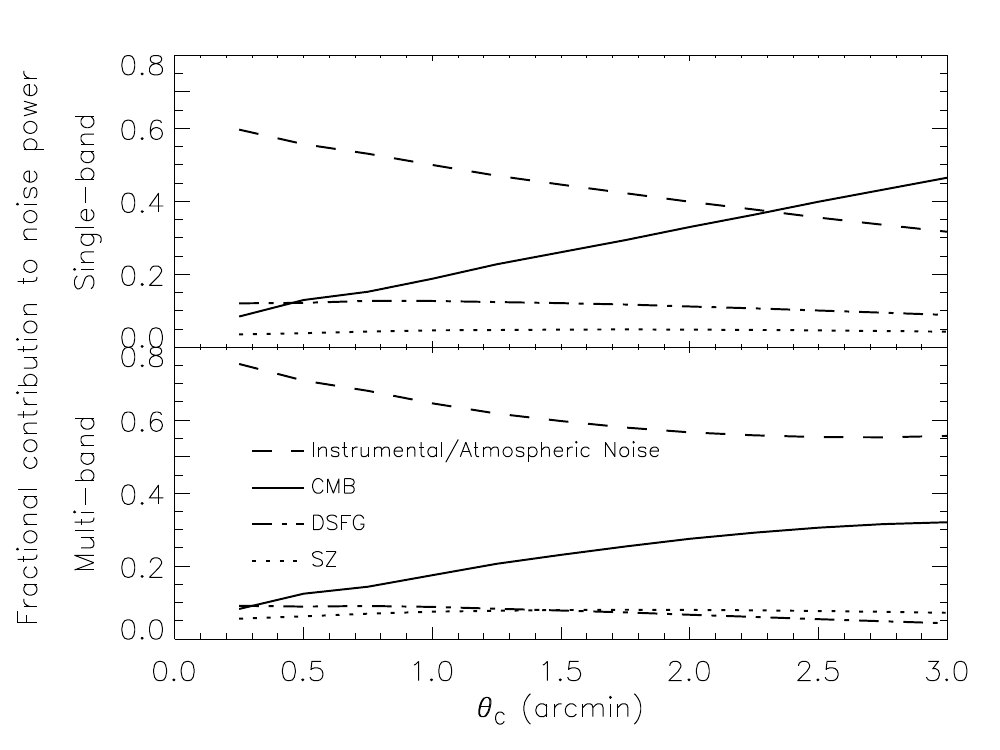}
\caption{The expected fractional contribution to noise power in the filtered maps for
the single-band SPT survey of \citet{Vanderlinde:10} (top panel)
and the multi-frequency SPT survey \citep[e.g., ][bottom panel]{Reichardt:12}, plotted against the core size $\theta_c$ of the $\beta$-model profile used to build the filter. The analysis presented in \S 4 assumes $\theta_c$ = 0.5'.}
\label{filter_noise}
\vskip 8pt
\end{figure}

\begin{figure}[h]
\centering
\epsscale{1.2}
\plotone{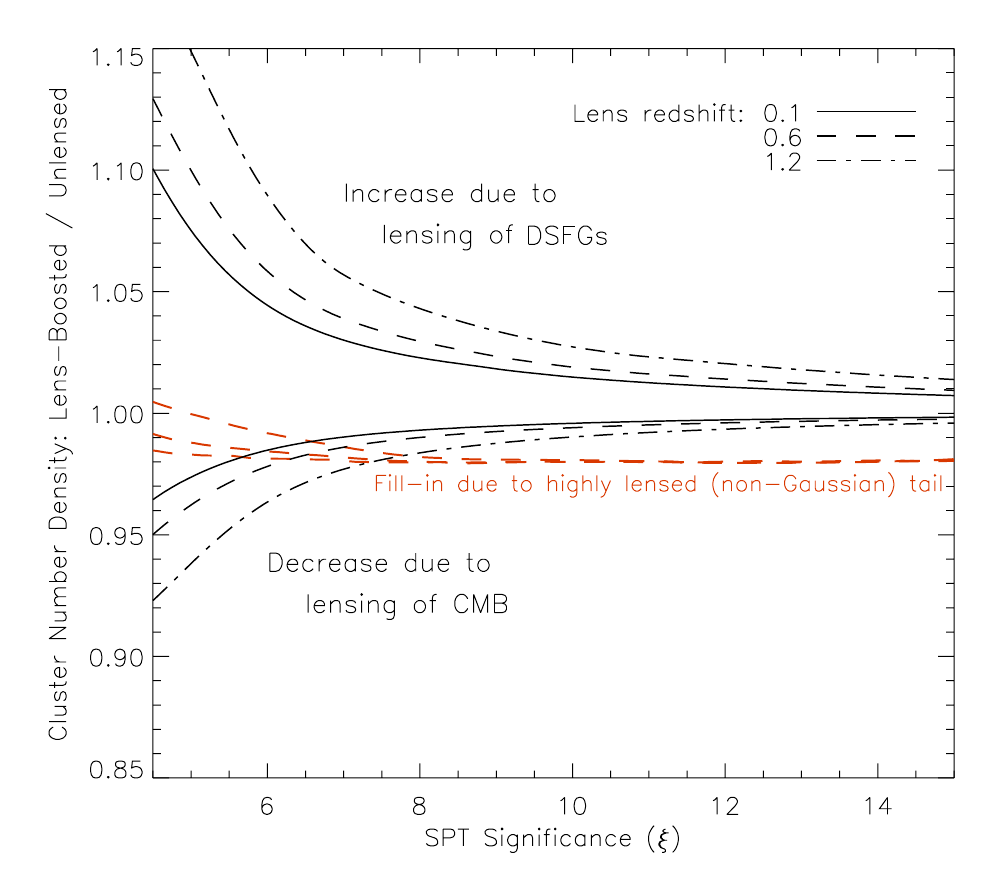}
\caption{The change in number counts of detected clusters due to lensing of the background DSFGs (greater than 1)
	and CMB (less than 1) at various redshifts and detection significances in a 150GHz single-band SPT cluster survey.}
\label{numdens_ratio}
\end{figure}

The expected contributions to noise in matched-filtered maps are shown 
in Figure \ref{filter_noise} for both the single-band SPT cluster survey of \citet{Vanderlinde:10}
(the example we are using) and the multi-frequency SPT survey \citep{Reichardt:12}. 
The unlensed DSFG background contributes
$12\%$ of the noise power in the single-frequency 
matched-filtered map, while the CMB 
contributes somewhat more, depending on the filter scale.
At each point in $\xi-$z space, we interpolate an effective new level for these contributions
based on the ratio of lensed-to-unlensed matched-filtered background
distributions (as shown in Fig \ref{sigma_ratio}).
We convolve this new noise level with the mass function, and compare the result
to the original, effectively isolating the effect of lensing of the backgrounds
on the observed number count of galaxy clusters.
The ratio of these number density grids is shown by the black curves
in Figure \ref{numdens_ratio}, for a variety of redshifts and significance levels.

We find that the lensing of the DSFGs increases their effective noise contribution,
increasing the Eddington bias and preferentially making clusters easier  to detect.
This results in an increase in the number density of 
detected clusters. Lensing of the CMB has the opposite effect, decreasing
the effective noise at cluster locations, making it less likely  that
a cluster would scatter up into a flux-limited sample.

Finally, we explore the effect of convolving the best-fit log-logistic 
distributions in place of the Gaussian fits for the DSFG noise contribution.
We interpolate the $\alpha$ and $\beta$ parameters (Eq. \ref{eq:loglogistic}) to each position in the $\xi-$z grid,
calculate the appropriate log-logistic, and convolve it alongside the Gaussian contributions
from other sources (unlensed CMB, instrumental and atmospheric).
Taking the ratio of the resulting grid with the original, we find a similar set of curves as above,
with a slight reduction due to the tail of highly magnified lensed point sources, which effectively ``fill in''
the 150 GHz SZ decrement. We difference the two sets of lensing-boosted number densities to estimate the
rate of such strong-lensing erasures, and find they occur at roughly the $2\%$ level, as shown
by the red dashed curves in Figure \ref{numdens_ratio}.

We note that while these effects are qualitatively robust
against the details of the survey being considered,
the exact magnitude of each will depend on the specific properties of the survey,
such as frequency coverage, resolution, and noise.
Adding the 90 GHz channel of SPT \citep[as in ][]{Reichardt:12}
reduces the DSFG contribution to the matched-filter noise (see Fig. \ref{filter_noise}),
such that when combined with the CMB contribution
the net effect of lensing on number counts is near zero.

\section{Discussion and Conclusion}
\label{sec:discussion}

A single-band cluster survey like that of \citet{Vanderlinde:10} can be affected by strong lensing of
either DSFGs or the CMB at the level of roughly 10\% in the number counts.
However, these effects largely offset each other,
making strong lensing effects a small source of bias.
While this remains subdominant to Poisson errors in the final SPT catalog of $\sim700$ clusters
(and will be further reduced by the multiple frequency bands of the full SPT survey),
a strong redshift dependence could lead to a bias in the inferred growth function or
cosmological parameter values such as the dark energy equation of state.

The long tail of strongly-lensed background sources additionally leads to a $\sim2\%$
decrease in detected cluster numbers independent of redshift or the original detectability of clusters.
We interpret this to mean that in $\sim2\%$ of cases, depending only on the geometry of the source-lens-observer system, 
strong lensing of the background DSFGs will sufficiently obscure the SZ signal of the foreground cluster lens that the cluster would be missed.
Again, we find this to be a small effect relative to the Poisson noise in current 
catalogs. While this is a very dramatic effect when it happens
(e.g., SPT-CLJ2332-5358 overlaps with a strongly lensed galaxy that is obviously
obscuring some of the SZ decrement), it is extremely rare, and is at
most the third-largest strong lensing-induced concern (swamped by the 
increase or decrease in scatter caused by lensing of the DSFGs or the CMB).

These effects will become significant in future surveys such as CCAT or SPT-3G,
where catalogs containing of order 10,000 clusters are expected.
They can, however, be mitigated with multi-band data,
where the contribution of the CMB and DSFG background to the final noise in the SZ map is reduced.

\acknowledgements{We thank Alex van Engelen for pointing out the
importance of CMB lensing. This work was
supported by the Canadian Institute for Advanced Research, as well as
NSERC Discovery and the Canada Research Chairs program. YDH acknowledges the support of FQRNT through International Training Program and Doctoral Research scholarships. We thank the SPT team for numerous useful discussions.}

\bibliographystyle{apj}
\bibliography{references}

\end{document}